\documentstyle[fleqn,11pt,amsmath,amssymb,epsfig,color,times]{article}

\def\thecomma{\ifx,\thenext \else\ifx;\thenext \else\ifx.\thenext
  \else\ifx!\the next \else\ifx:\thenext \else \  \fi\fi\fi\fi\fi}
\def\condblank{\futurelet\thenext\thecomma}

\def\invivo{{\it in vivo}\thecomma}
\def\invitro{{\it in vitro}\thecomma}

\begin{document}
\begin{titlepage}

\mbox{} \vspace*{2.5\fill} {\Large\bf
\begin{center}
%
The Physics of Living Neural Networks
%
\end{center}
}

\renewcommand{\thefootnote}{\fnsymbol{footnote}}

\vspace{1\fill}

\begin{center}
{\large Jean-Pierre Eckmann$^{1}$, Ofer Feinerman$^{2}$, Leor Gruendlinger$^{3}$, Elisha
Moses$^{2,}$\footnote[2]{Corresponding author. E-mail address:
elisha.moses@weizmann.ac.il}, Jordi Soriano$^{2}$, Tsvi Tlusty$^{2}$}
\end{center}

\begin{small}
\begin{center}
$^{1}${\sl D\'epartement de Physique Th\'eorique and Section de Math\'ematiques,
Universit\'e de Gen\`eve. CH-1211, Geneva 4, Switzerland}
\\[2mm]
$^{2}${\sl Department of Physics of Complex Systems, Weizmann Institute of Science.
Rehovot 76100, Israel}
\\[2mm]
$^{3}${\sl Department of Neurobiology, Weizmann Institute of Science. Rehovot 76100,
Israel}
\\[2mm]
\end{center}
\end{small}

\kern 1cm \hrule \kern 3mm
\begin{small}
\noindent
{\bf Abstract}
\vspace{3mm}


Improvements in technique in conjunction with an evolution of the theoretical and
conceptual approach to neuronal networks provide a new perspective on living neurons in
culture. Organization and connectivity are being measured quantitatively along with other
physical quantities such as information, and are being related to function. In this
review we first discuss some of these advances, which enable elucidation of structural
aspects. We then discuss two recent experimental models that yield some conceptual
simplicity. A one--dimensional network enables precise quantitative comparison to
analytic models, for example of propagation and information transport. A two-dimensional
percolating network gives quantitative information on connectivity of cultured neurons.
The physical quantities that emerge as essential characteristics of the network \invitro
are propagation speeds, synaptic transmission, information creation and capacity.
Potential application to neuronal devices is discussed.

\kern 2mm

\noindent {\em PACS:}\  87.18.Sn, 87.19.La, 87.80.-y, 87.80.Xa, 64.60.Ak


\noindent {\em Keywords:}\  complex systems, neuroscience, neural networks, transport of
information, neural connectivity, percolation.

\end{small}
\kern 2mm \hrule \kern 1cm

\vfill

\end{titlepage}

\tableofcontents

\newpage


\section{Preamble}
\label{sec:Preamble}

The mysteries of biology are a challenge not only to biologists, but to a wide spectrum
of scientists. The role of both experimental and theoretically inclined physicists in
this quest is slowly shifting towards more conceptually driven research. Historically,
such efforts as the Phage group \cite{Phage-1992}, the discovery of DNA \cite{DNA-2003}
and the classical field of biophysics saw a melding of physicists into the biological
field, carrying over experimental techniques into biology. Recently, however, a new and
different approach has become possible for physicists working in biology, with the
novelty that it includes equal contributions from theoretical and experimental
physicists.

There are two main changes of paradigm that distinguish this approach, giving hope for
the creation of a profound impact. First is the realization that biology is characterized
by the interaction of a large number of independent agents, linked and tangled together
in a functional network. The properties of such networking systems can be elucidated by
ideas of statistical physics and of dynamical systems, two fields that have evolved and
now merge to treat the concept of complexity. These fields are now mature enough to
provide a qualitative and quantitative understanding of the most complex systems we know,
those of life and of the brain. Examples for the success of the statistical physics /
dynamical systems approach range from genetic, computer and social networks, through the
intricacies of cell motility and elasticity, all the way to bio--informatics, and its
results already have a wide range of consequences.

Second is the insistence of physicists to set conceptual questions at the forefront, and
to look for universally applicable questions, and, hopefully, answers.

The mere complexity of biology, and its apparently unorganized, slowly evolving nature,
seem to prevent any conceptual approach. But conceptual ideas will eventually, so we
hope, lead to a deeper understanding of what biology is about.

In Molecular Cell Biology, it is the concept of creating life, making an artificial cell
that has locomotion and the ability to divide. For example, a physicist might ask what
physical limitations make $130^{\circ}$C the highest temperature at which life is still
sustainable, and the conceptual answer might have relevance for astrobiology.

In this review, we present a summary of how the general ideas and principles described
above can be applied to studies of neural networks grown from living neurons. The
conceptual problem involved is currently still unclear, and part of our goal in this
review is to frame the right questions and to outline the beginning of some answers. We
will need to use the ideas of statistical mechanics, of graph theory and of networks. But
we will be aiming at a slightly more complex and higher level of description, one that
can also describe the computation that is going on in the network. In the background is a
final goal of comparison to processes of the real brain.

This quest should not be confused with the well--developed subject of neural networks. In
that case, the general question seems to be whether a network with memories and certain
types of connections can be taught to do certain tasks or to recognize certain patterns.

Another important field is that, more akin to engineering, of designing new forms of
computation devices made of or on living neural substrates. These have wide ranging
conceptual and applicative consequences, from possible new computers to the newly
emerging reality of neural implants that can solve medical infirmities such as blindness.

Here, we concentrate on the more modest, but more conceptual, issue of computation
occurring in a network of neurons \invitro. This seemed to us at the same time simple
enough so that we are able to actually do experiments and still sufficiently biological
to carry some of the characteristics of living matter.

\section{Introduction}
\label{sec:Intro}

Living neural networks grown \invitro show network behavior that is decidedly different
from that of any part in the brain and from any neural network in the living animal. This
has led neurobiologists to treat with a modicum of doubt the applicability to the brain
of conclusions stemming from {\it \invitro} networks. Many neurobiologist turn instead to
the brain slice, which has many of the advantages of \invitro cultures---e.g., ease of
access for microscopy, electrodes and chemicals---but has the correct (``organotypic")
structure.

For a physicist, this is not a disadvantage, but rather an opportunity to ask why a
neural culture is less capable of computation than the same neurons grown in the brain.
We are going, in fact, to ask what are the capabilities of the culture, how they fall in
quality from those of neurons grown in the brain, and what causes this disability.
Information that is input from the noisy environment of the external world is processed
by the culture in some way, and creates an internal picture, or representation of the
world. We will be asking what kind of representation does the {\it \invitro} culture make
of the external world when input is injected to it from the outside.

It turns out, not too surprisingly, that the representation that the culture creates is
simplistic. The repertoire of responses that the network is capable of making is limited.
On the other hand, we will see that the neurons do communicate; they send signals that
are received and processed by other neurons, and in that sense we are in the presence of
a network of identical units that are connected together in an almost random way.

One thing that we have kept in mind throughout these studies is the observation that
these units are designed by Nature to make connections, and since they have no
pre-assigned task, like ``talking'' or ``seeing'' that would determine their input, they
learn to interact only through their genetic program and external stimuli. They have thus
no particular aim other than connecting and communicating, and it is precisely this
neutrality of purpose that makes their study ideal for a precise quantitative
investigation.

We will furthermore see that the dimensionality of the network impacts strongly on its
connectivity, and therefore plays an important role for its possible behavior, and we
will show how one--dimensional and two--dimensional networks can be understood in terms
of these concepts. The connectivity of a full network has not been revealed to date by
any other means, and its unveiling can have many implications for neurobiology.

\section{Characterization of the \invitro neural network}

In this review we present some new experimental designs in which simple geometries of the
culture and the mode of excitation allow for precise measurements of network activity.
The simple geometry and controlled experiments allow the comparison with detailed
theoretical models. The excellent agreement gives confidence that the concepts used in
the models are applicable to the culture. There arises a picture of some kind of
self--organization, as a system described by simple, randomly organized connectivity.

We will see that a number of conceptual models describe the behavior of the culture in a
precise manner. Propagation speeds in uni--dimensional cultures can be accurately
described by a model of Osan and Ermentrout \cite{Osan-2002} that is based on a
continuous integrate and fire (IF) model. Looking at information transport in these 1D
cultures shows that a simple model of concatenated Gaussian information channels (the
``Gaussian chain'') describes the decay of information with distance extremely well
\cite{Feinerman-2005,Feinerman-2006}. In two dimensions, connectivity is well described
by a percolation model, describing a random, local network with Gaussian degree
distribution \cite{Breskin-2006}.

The models for describing the culture involve simple rules of connection, for example
those leading to a linear chain of simple processing units. This defines both the
information transport capability and the wave propagation speeds. Since the models
reproduce well the experimental results, one may conclude that it is indeed the simple,
random connectivity of the neural cultures that limits their computing possibilities.

At this point a comparison to the brain becomes more tangible. Obviously the brain will
be different in having a blueprint for connectivity, not leaving the details of
connection to chance. The brain is three dimensional, much more complex and ramified, and
if it were left to chance how the connections are made it would be completely
unstructured. The basic difference is that the brain does not leave its connectivity to
chance, or to random processes. Connections are presumably determined in the brain
according to functionality, with the view of enabling specific functions and processes. A
full contingent of signaling chemicals is used to guide axons and neurons as they locate
themselves within the network. In the absence of such design the neurons of \invitro
networks connect as best as they can, seeking out chemical signals and cues. All they
find, however, is whatever chemicals nearby neurons emit, carried by random diffusion and
advection drifts in the fluid above the culture.

Neural network activity integrates effects that stem from both the single neuron and the
network scales. Neuronal cultures provide a major tool in which single neurons may be
singled out to study their properties, as well as pair--wise interaction. However, the
connectivity of the cultured network
\cite{Breskin-2006,Nakanishi-1998,Nakanishi-1999,Nakanishi-2000} is very different from
\invivo structures. Cultured neurons are thus considered less ideal for studying the
larger, network scales.

From a physicist's point of view the dissimilarities between cultured and \invivo network
structures are less disturbing. On the contrary, connectivity and network properties can
be regarded as an experimental handle into the neuronal culture by which
structure-function relations in the neural network can be probed. The success of this
direction relies on two main factors. The first is that the structure-function relations
are indeed easier to measure and to express in the context of simplified connectivity
patterns. The second, and perhaps more intricate, point is that such relations could then
be generalized to help understanding activity in the complex and realistic \invivo
structures.

\begin{figure*}[!ht]
\begin{center}
\includegraphics[width=16cm]{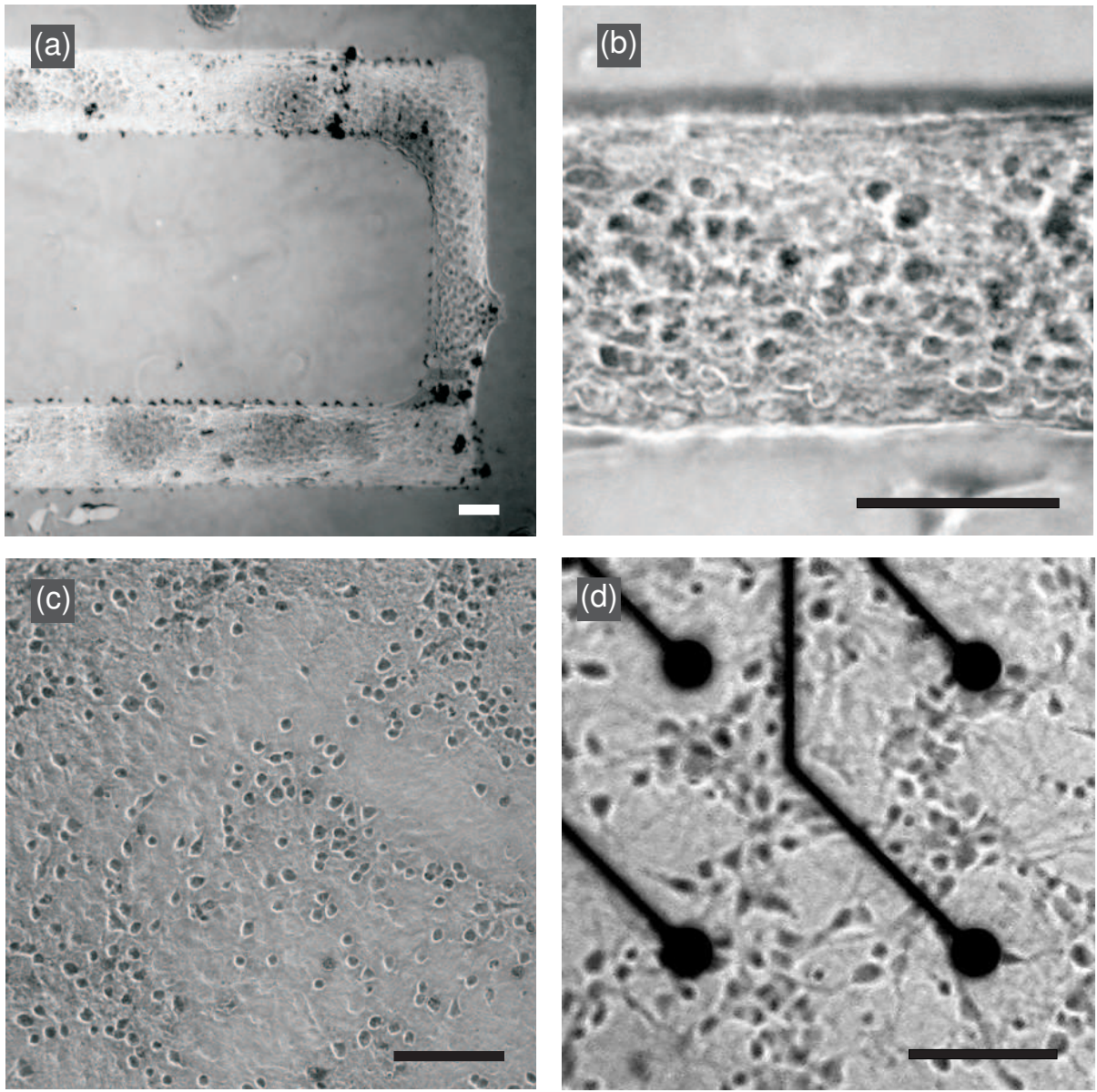} \vspace{0.0cm}
\caption{Examples of neural cultures. Dark spots are neurons. (a) Neurons plated in a 1D
culture on a pre--patterned line $170$ $\mu$m wide. (b) Detail of the neurons on the line
in (a). (c) Neurons plated in a 2D culture on glass coverslips and, (d) multielectrode
arrays (MEA), showing the electrodes and the neurons in the area nearby. Scale bars are
$100$ $\mu$m.} \label{Fig:Cultures}
\end{center}
\end{figure*}

\section{Cell culturing}\label{sec:Cultures}

The history of cell culturing goes back to the end of the 19th and beginning of the 20th
century, when Roux and then Harrison showed that cells and neurons can be maintained
alive outside of the animal. However, it was not before the 1950's that immortalization
of the cell was achieved by causing a cancer--like bypass of the limitation on the number
of its divisions. This enabled the production of cell ``lines", and cultures evolved to
become a widespread and accepted research tool. Immortalization was to a large extent
irrelevant for neurons, since they do not divide, and reliance on a ``primary'' culture
of neurons extracted directly from the brain remained the norm.

The currently accepted protocol for culturing primary cultures of neurons is well defined
and usually involves the dissection of brains from young rats, either embryonic or
neo--natal. Cultures are typically prepared from neurons of specific regions of the
brain, such as the hippocampus or cortex, dissociated, and plated over glass\footnote{The
plating consists of coating glass coverslips with a thin layer of adhesion proteins.
Neurons, together with nutrients, are placed homogeneously over the glass, adhering to
the proteins.}. Neurons start to develop connections within hours and already show
activity $3-4$ days after plating \cite{Wagenaar-2006}. Experiments are normally carried
out at day $14-21$, when the network is fully mature. A complete and detailed description
of the culturing process can be found for example in
\cite{Papa-1995,Murphy-1996,Banker-1998}. The culturing process is reproducible and
versatile enough to permit the study of neural cultures in a variety of geometries or
configurations, and with a broad spectrum of experimental tools (Fig.\
\ref{Fig:Cultures}). Modern techniques permit to maintain neural cultures healthy for
several months \cite{Gopal-1996,Potter-2001}, making them excellent model systems to
study development \cite{Wagenaar-2006,Marom-2002}, adaptation
\cite{Fuhrman-2002,Giugliano-2004}, and long--term learning
\cite{Potter-2001,Marom-2002,Shahaf-2001,DeMarse-2001,EytanMarom-2003} and plasticity
\cite{Maeda-1998,Jimbo-1999}.

\section{Experimental approaches}
\label{sec:Experimental-app}

In this section we review several experimental approaches to study neural cultures,
emphasizing those techniques that have been used in our research. We start with the now
classical patch--clamp technique, and then describe novel techniques that are adapted to
questions which arise naturally when biological systems are considered from a physicists
point of view. Such experimental systems are specifically capable of supplying
information on the activity of many neurons at once. We give here a brief summary of the
advantages and limitations of each of these techniques.

\subsection{Patch-clamp}

Patch--clamp is a technique that allows the recording of single ion--channel currents or
the related changes in cells' membrane potentials \cite{Hamill-1981,Neher-1992}. The
experimental procedure consists of attaching a micropipette to a single cell membrane,
and it is also possible to put the micropipette in contact with the intracellular medium.
A metal electrode placed inside the micropipette reads any small changes in current or
voltage in the cell. Data is then amplified and processed with electronics.

The advantage and interest of the patch--clamp technique is that it allows to carry out
accurate measurements of voltage changes in the neurons under different physiological
conditions. Patch--clamp is one of the major tools for modern neuroscience and drug
research \cite{Owen-2002,Stett-2003a}, and it is in a continuous state of improvement and
development. Current techniques allow to study up to $12$ neurons simultaneously, which
is an impressive achievement given the difficulty of the accurate placement of the
micropipettes and the measurement of the quite weak electrical signals. Remarkable
progress has been attained with the development of chip--based patch--clamp techniques
\cite{Ionescu-2005}, where silicon micromachining or micromolding is used instead of
glass pipettes. In general, however, given the sophistication of the equipment that
patch--clamp requires, measurement of substantially larger number of neurons are not
feasible at this point.

As we will see below, techniques addressing many more neurons exist, but they do not
reach the precision of the patch--clamp technique.

\subsection{Fluorescence and calcium imaging}\label{Subsec:calcium}

Fluorescence imaging using fluorescent dyes can effectively measure the change in
electric potential on the membrane of a firing neuron (``voltage sensitive dyes"), or the
calcium increase that occurs as a consequence (``calcium imaging")
\cite{Kao-1994,Gee-2000}. After incubation with the fluorescent dye, it is possible to
measure for a few hours in a fluorescent microscope the activity in the whole field of
view of the objective. In our experiments \cite{Breskin-2006}, this sample can cover on
the order of 600 neurons. Since the culture typically includes $10^5$ neurons, this
sample represents the square root of the whole ensemble, and should give a good
estimation for the statistics of the network.

The response of calcium imaging fluorescence is typically a fast  rise (on the order of a
millisecond) once the neuron fires, followed by a much slower decay (typically a second)
of the signal. This is because the influx of calcium occurs through rapidly responding
voltage gated ion channels, while a slower pumping process governs the outflow. This
means that the first spike in a train can be recognized easily, but the subsequent ones
may be harder to identify. In our measurements we find that the bursting activity
characteristic of the network rarely invokes more than five spikes in each neuron, and
within this range the fluorescence intensity is more or less proportional to the number
of spikes.

The advantages of the fluorescence technique are the ease of application and the
possibility of using imaging for the detection of activity, the fast response and the
large field of view. It also benefits from continuous advances in imaging, such as the
two--photon microscopy \cite{Garaschuk-2000}, which substantially increases depth
penetration and sensitivity, allowing the simultaneous monitoring of thousands of
neurons. The disadvantages are that the neurons are chemically affected, and after a few
hours of measurement (typically $4-6$) they will slowly lose their activity and will
eventually die. Subtle changes in their firing behavior may also occur as a result of the
chemical intervention.

\subsection{Multielectrode arrays (MEA)}

Placing numerous metal electrodes on the substrate on which the neurons grow is a natural
extension of the single electrode measurement. Two directions have evolved, with very
different philosophies. An effort pioneered by the Pine group
\cite{Pine-1980,Regehr-1989,Maher-1999}, and represented by the strong effort of the
Fromherz group \cite{Zeck-2001,Fromherz-2003,Fromherz-2005}, places specific single
neurons on top of a measuring electrode. Since neurons tend to connect with each other,
much effort is devoted to keep them on the electrode where their activity can be
measured, for example by building a ``cage" that fences the neurons in. This allows very
accurate and precise measurements of the activity, but is highly demanding and allows
only a limited number of neurons to be measured.

The second approach lays down an array of electrodes (MEA) on the glass substrate, and
the neurons are then seeded on the glass. Neurons do not necessarily attach on the
electrodes, they are rather randomly located in various proximities to the electrodes
(Fig.\ \ref{Fig:Cultures}d). The electrode will in general pick up the signal of a number
of neurons, and some spike sorting is needed to separate out activity from the different
neurons. The different relative distance of the neurons to the electrodes creates
different electric signatures at the electrodes allowing efficient sorting, usually
implemented by a simple clustering algorithm.

This approach, originating with the work of Gross \cite{Gross-1979,Gross-1993}, has
developed into a sophisticated, commercially available technology with ready-made
electrode arrays of different sizes and full electronic access and amplification
equipment \cite{Potter-2001,Jimbo-1992,Meister-1994,Potter-2006}. Electrode arrays
typically include $64$ electrodes, some of which can also be used to stimulate the
culture. Spacing between the electrodes can vary, but is generally on the order of a few
hundred $\mu$m's.

The advantages of the MEA are the precise measurement of the electric signal, the fast
response and high temporal resolution. A huge amount of data is created in a short time,
and with sophisticated analysis programs, a very detailed picture can be obtained on the
activity of the network. This technique is thus extensively used to study a wide range of
problems in neuroscience, from network development to learning and memory
\cite{Wagenaar-2006,Potter-2001,Fuhrman-2002,Giugliano-2004, Shahaf-2001,DeMarse-2001,
EytanMarom-2003,Maeda-1998,Jimbo-1999} and drug development
\cite{Morefield-2000,Stett-2003b}. The disadvantages of the MEA are that the neurons are
dispersed randomly, and some electrodes may not cover much activity. The measured
extracellular potential signal is low, on the order of several $\mu$V, and if some
neurons are located at marginal distances from the electrode, their spikes may be
measured correctly at some times and masked by the noise at others.

The lack of accurate matching between the position of the neurons and the one of the
microelectrodes may be critical in those experiments where the identification of the
spiking neurons is important. Hence, new techniques have been introduced in the last
years to enhance the neuron--to--electrode interfacing. Some of the most important
techniques are the neural growth guidance and cell immobilization
\cite{Maher-1999,Zeck-2001,Fromherz-1991,Sanjana-2004}. These techniques, however, have
proven challenging and are still far from being standard.

MEA is also limited to the study of the response of single neurons, and cannot deal with
collective cell behavior. Phenomena that take place at large scale and involve a huge
amount of neurons, such as network structure and connectivity, can not be studied in
depth with MEA techniques. This motivated the development of new techniques to {\it
simultaneously} excite a large region of a neural network and study its behavior.
However, methods for excitation of a large number of neurons often lack the recording
capabilities of MEA. In addition, these techniques normally use calcium imaging to detect
neuronal activity, which shortens drastically the duration of the experiments from days
to hours.

An innovative technique, in the middle between MEA and collective excitation, consists on
the use of light--directed electrical stimulation of neurons cultured on silicon wafers
\cite{Artem-2004,Colicos-2001}. The combination of an electric current applied to the
silicon surface together with a laser pulse creates a transient ``electrode" at a
particular location on the silicon surface, and by redirecting the location of the laser
pulse it is possible to create complex spatiotemporal patterns in the culture.

\subsection{Collective stimulation through bath electrodes}

Collective stimulation can be achieved by either electric or magnetic fields applied to
the whole culture. Magnetic stimulation techniques for neural cultures are still under
development, although they have been successfully applied to locally excite neurons in
the brain \cite{Hallett-2000,Pascual-2000}. Electrical stimulation is becoming a more
common technique, particularly thanks to its relative low cost and simplicity. Reiher
{\it et al.} \cite{Reiher-2005} introduced a planar Ti--Au--electrode interface
consisting on a pair of Ti--Au electrodes deposited on glass coverslips and separated by
$0.3$ mm. Neurons were plated on the coverslips and were collectively stimulated through
the electrodes. Neural activity was measured through calcium--imaging.

Our experiments use both chemical \cite{Feinerman-2005} and electric stimulation
\cite{Breskin-2006}. The electric version is a variation of the above described
technique. Neurons are excited by a global electrical stimulation applied to the entire
network through a pair of Pt--bath--electrodes separated by $15$ mm, with the coverslip
containing the neural culture centered between them. Neural activity is measured with
calcium--imaging.

The major advantage of collective stimulation is that it permits to simultaneously excite
and study the response of a large neural population, on the order of $600$ neurons in our
experiments \cite{Breskin-2006}. The major disadvantage is that calcium--imaging limits
the duration of the experiments, and that repeated excitations at very high voltages
($\gtrsim 10$ V) significantly damage the cells and modify their behavior
\cite{Reiher-2005}.


\section{The activity repertoire of neuronal cultures}

When successfully kept alive in culture, neurons start to release signaling molecules,
called {\it neurotransmitters}, into the environment around them, supposedly intended to
attract connections from their neighbors \cite{Kandel-1995,BenAri2006}. They also grow
long, thin extensions called {\it axons} and {\it dendrites} that communicate with these
neighbors, in order to transmit (axons) or receive (dendrites) chemical messages. At the
meeting point between an axon and a dendrite there is a tiny gap, about $200-400$ nm
wide, called a chemical {\it synapse}. At the dendritic (receiving) side of the synaptic
gap there are specialized receptors that can bind the neurotransmitters released from the
axon and pass the chemical signal to the other side of the gap. The effect of the message
can be either {\it excitatory}, meaning that it activates the neuron that receives the
signal, {\it inhibitory}, i.e., it de--activates the target neuron, or {\it modulatory},
in which case the effect is usually more prolonged and complex. The release of
neurotransmitters into the synapse is usually effected by a short ($\approx 3$ msec)
electrical pulse, called {\it action potential} or {\it spike}, that starts from the body
of the neuron and passes throughout the axon. The neuron is then said to have {\it fired
a spike}, whose effect on neighboring neurons is determined by the strengths of the
synapses that couple them together.

\begin{figure}[hbt]
\begin{center}
\includegraphics[width=12cm]{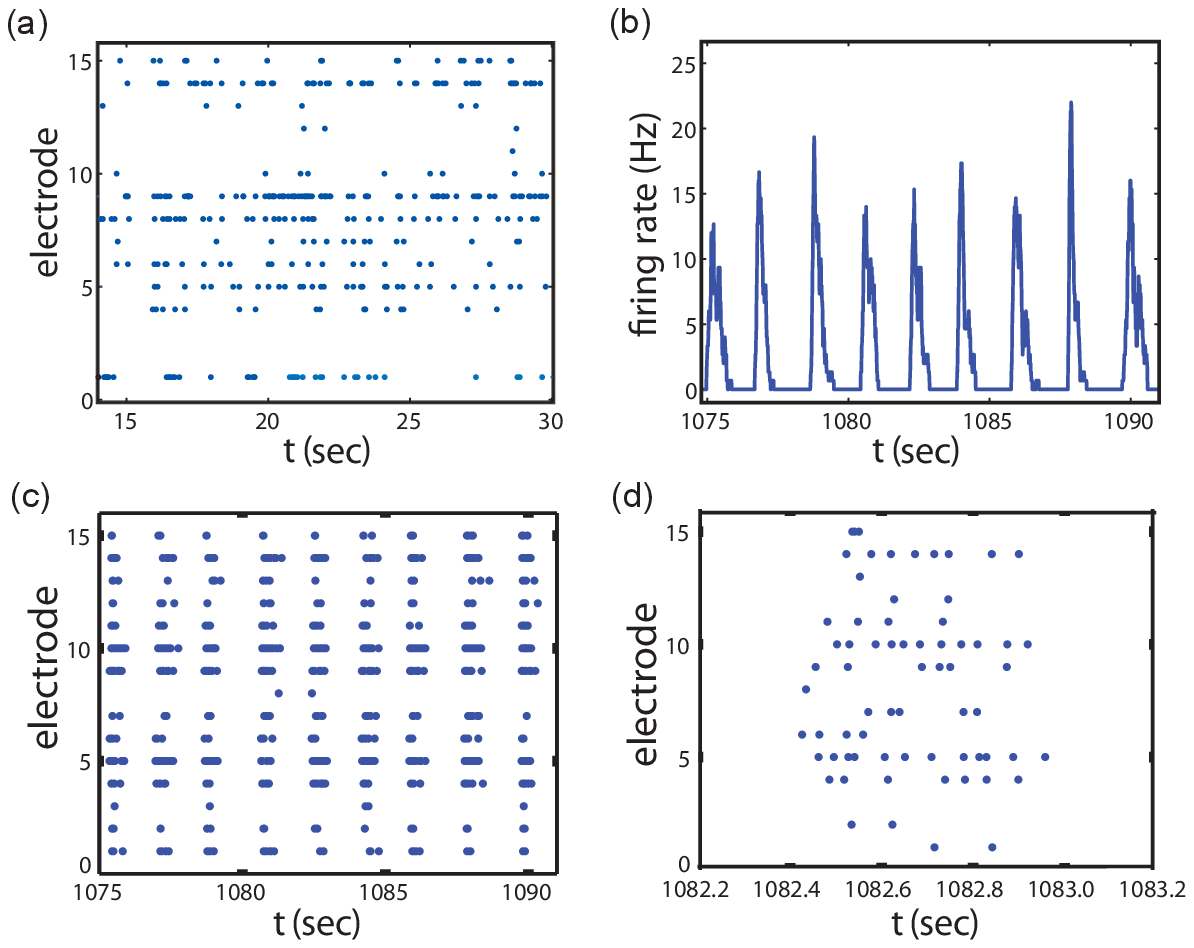}
\caption {Activity repertoire of neuronal cultures: (a) Raster plot of asynchronous
firing, showing time on the x axis and electrode number on the y axis. (b) Network burst,
showing the average firing rate across the culture as a function of time. (c) Raster plot
of burst activity. Bursts appear as simultaneous activity in nearly all electrodes,
separated by silent periods. (d) Zooming in one burst in (c) shows that there is an
internal spiking dynamics inside the burst. Data taken from a multielectrode array
recording by E. Cohen from hippocampal cultures at the M. Segal lab.}
\label{fig:EyalBurst}
\end{center}
\end{figure}

As the cultured neurons grow and connect with each other, they form a network of coupled
cells, whose firing activity usually takes one of three main forms: (i) Asynchronous
firing (Fig.~\ref{fig:EyalBurst}a). Neurons fire spikes or bursts in an uncoordinated
manner. This typically occurs at very early developmental stages (e.g., $2-5$ days
\invitro) or in special recording media \cite{LathamExp}. (ii)  Network bursts (Figs.
\ref{fig:EyalBurst}b-d). This is by far the most common pattern of activity, where short
bursts of intense activity are separated by long periods of near--quiescence. (iii)
Seizure--like activity. This is an epilepsy--like phenomenon, characterized by very long
(tens of seconds) episodes of intense synchronized firing, that are not observed in
neuronal cultures under standard growth conditions \cite{DeLorenzo}.

\subsection{Network bursts}
Fig.~ \ref{fig:EyalBurst} shows an example of a network burst, which is a prominent
property of many neuronal cultures \invitro \cite{Buzsaki2004,Corner}. Several days after
plating, cultured neurons originating from various tissues from the central nervous
system display an activity pattern of short bursts of activity, separated by long periods
of near--quiescence (called inter--burst--interval). This kind of activity persists as
long as the culture is maintained \cite{Corner,CornerShort,vanPeltLongTerm}.

Network bursts also occur in organotypic slices grown \invitro \cite{BenAriTins,Streit}.
Similar patterns of spontaneous activity were found \invivo in brains of developing
embryos \cite{BenAriTins} and in the cortex of deeply anesthetized animals
\cite{Diamond}. Recent evidence suggests that they may also occur in the hippocampus of
awake rats during rest periods after intensive exploration activity \cite{ExploringRats}
(called ``sharp waves''), as well as during non--REM sleep \cite{Battaglia-2004}. Network
bursting has been produced experimentally already in the 50's in cortical ``slabs''
\invivo (small areas of the cortex, whose neuronal connections with the rest of the brain
are cut, while blood supply is maintained intact \cite{Burns,TimofeevSlabs}). After the
surgery the animals recover, but within a few days the slab typically develops network
bursts.

These varied physiological states are all characterized by the absence, or a prolonged
reduction, in sensory input. Recent modeling and experimental work suggests that low
levels of input may indeed be a requirement for network bursting
\cite{LathamExp,LathamModel,Houweling,Potter}, along with a strong--enough coupling
between the neurons \cite{AlexMisha}. Interestingly, bursting appears to be much more
sensitive to the background level of input than to the degree of coupling. During periods
of low activity, neurons seem to accumulate a large pool of excitable resources. When a
large enough group of these ``loaded'' neurons happens to fire synchronously, they start
a fast positive--feedback loop that excites their neighbors and can potentially activate
nearly all the neurons in the culture within a short and intense period. Subsequently,
the neurons start a long process of recovering the depleted resources until the next
burst is triggered.

Modeling studies suggest that when the input to the neurons is strong enough, significant
sporadic firing activity occurs during the inter--burst phase and depletes the pool of
excitable resources needed to ignite a burst, and thus the neurons fire asynchronously.
On the other hand, prolonged input deprivation (e.g., in cortical slabs immediately
post-surgery) appears to cause a homeostatic process in which neurons gradually increase
their excitability and mutual coupling. Both changes increase the probability of firing
action potentials, but in different ways: Increasing excitability can cause neurons to be
spontaneously active independently of their neighbors, i.e., in an asynchronous manner,
while stronger coupling promotes synchronization. For an unknown reason, under
physiological conditions the mutual coupling seems to be increased much more than the
individual excitability, and after a few days the neuronal tissue starts bursting.

\subsection{Epileptiform activity}

In addition to having a yet--unknown physiological significance, network bursts may
potentially serve as a toy model for studying mechanisms of synchronization in epilepsy.
One of the \invitro models for epilepsy is seizure-like activity, a phenomenon
characterized by very long (tens of seconds) periods of high--frequency firing. It is
typically observed either in whole brains \cite{SteriadeReview}, in very large cortical
slabs \cite{TimofeevSlabs}, or when using certain pharmacological protocols \invitro
\cite{DeLorenzo}. While network bursts are much shorter than seizure--like activity, some
experimental and theoretical evidences indicate that when the density and size of a
neuronal ensemble is increased, network bursts are replaced by prolonged seizure--like
activity \cite{TimofeevSlabs,Houweling}. The question of why neuronal ensemble size
affects bursting is still being debated, and non-synaptic interactions in the
densely--packed tissue of neurons and glia in the brain possibly play an important role,
e.g., through regulation of the extracellular $K^+$ concentration
\cite{Feng2006,AstroEpilepsy}.

\subsection{MEA and Learning}

In the context of cultures, one can speak of learning in the sense of a persistent change
in the way neuronal activity reacts to a certain external stimulus. It is widely believed
that the strength of the coupling between two neurons changes according to the intensity
and relative timing of their firing. For a physicist, this suggests that neuronal
preparations have the intriguing property that the coupling term used in neuronal
population modeling is often not constant, but rather changes according to the activity
of the coupled systems. The correlation between synchronized firing and changes in
coupling strength is known in neurobiology as Hebb's rule \cite{Hebb,HenryMisha}.

The typical experiment consists of electrically stimulating groups of neurons with
various patterns of ``simulated action potentials" and observing the electrical and
morphological changes in the cultured neural networks, sometimes called ``neuronal
plasticity''. These cellular and network modifications may indicate how neurons in living
brains change when we learn something new. They may involve changes in the electrical or
chemical properties of neurons, in synapse number or size, outgrowth or pruning of
dendritic and axonal arbors, formation of dendritic spines, or perhaps even interactions
with glial cells.

An interesting approach worth noting is that of Shahaf and Marom \cite{Shahaf-2001}, who
stimulated the network through a single MEA electrode (the input), and measured the
response at a specific, different site on the electrode array (the output) within a short
time window after stimulation. If the response rate of the output electrode satisfied
some predetermined requirement, stimulation was stopped for a while, and otherwise
another round of stimulation was initiated. With time, the network ``learned" to respond
to the stimulus by activating the output electrode. One interpretation of these results
is that each stimulation, as it causes a reverberating wave of activity through the
network, changes the strengths of the coupling between the neurons. Thus, the coupling
keeps changing until the network happens to react in the desired way, at which point
stimulation stops and hence the coupling strengths stop changing.

It is perhaps interesting to note that with time, the network seems to ``forget'' what it
learned: spontaneous activity (network bursts that occur without prior stimulation) is
believed to cause further modifications in the coupling strengths and a new wiring
pattern is formed, a phenomenon reproduced by recent modeling work \cite{Chao-2005}.

Other studies have tried to couple neuronal cultures to external devices, creating hybrid
systems that are capable of interacting with the external world
\cite{DeMarse-2001,Bakkum-2004,Ruaro-2005}, for instance to control the movement of a
robotic arm \cite{Bakkum-2004}. In these systems, input is usually applied through
electrodes, and neurons send an output signal in the form of a network burst. Still, more
modeling work is needed before it would be possible to use bursts as intelligent control
signals for real--world applications. Even though they are so commonly seen and
intuitively understood, a true model that can, for instance, forecast the timing of
network bursts in advance is still not available.


\section{1-D neural cultures}

Since we are interested in studying cultures with many neurons, it is natural to look for
simplifying experimental designs. Such a simplification can be obtained by constraining
the layout of the neurons in the coverslip. Here we focus on the simplest topology of
uni--dimensional networks \cite{Feinerman-2005,Maeda-1995,Chang-2001,Segev-2002}, while
in the next section we address two--dimensional configurations.

The main advantage of the $1$--D architecture is that, to a first approximation, the
probability of two neurons to make a functional connection depends only on a single
parameter, namely their distance. A further simplification is introduced by using long
lines, since then the connection probability---sometimes referred to as {\it connectivity
footprint} \cite{Golomb-1997}---falls to zero on a scale much shorter than the length of
the culture, and thus it may be considered local. These two simplifications allow one to
obtain interesting results even though the study of one--dimensional neural systems is a
relatively young field.

Uni--dimensional networks may be regarded as information transmission lines. In this
review, we focus on two types of results, the propagation of information and the study of
the speed of propagating fronts. In these cases there is good agreement between theory
and experiments, which establishes for the first time measurable comparisons between
models of collective dynamics and the actual measured behavior of neural cultures.

\subsection{Experimental technique}

The idea of plating dissociated neurons in confined, quasi one--dimensional patterns was
introduced by Maeda {\it et al.} \cite{Maeda-1995}. Chang {\it et al.} \cite{Chang-2001}
and Segev {\it et al.} \cite{Segev-2002} used pre--patterned lines and photolithographic
techniques to study the behavior of $1$--D neural networks with MEA recording. Here we
focus on a new technique recently introduced by Feinerman {\it et al.}
\cite{Feinerman-2005,Feinerman-2006}. It consists of plating neurons on pre--patterned
coverslips, where only designated lines (usually $170$ $\mu$m wide and up to $8$ cm long)
are amenable to cell adhesion. The neurons adhere to the lines and develop processes
(axons and dendrites) that are constrained to the patterned line and align along it to
form functional connections. The thickness of the line is larger than the neuronal body
(soma), allowing a few cells to be located along this small dimension. This does not
prevent the network from being one--dimensional since the probability of two cells to
connect depends only on the distance between them along the line (but not on direction).
On average, a neuron will connect only to neurons that are located no more than $300-400$
$\mu$m from it. Neuronal activity is measured using fluorescence calcium imaging, as
described in Sec.\ \ref{Subsec:calcium}.

\subsection{Transmission of information}

Activity patterns are used by brains to represent the surroundings and react to them. The
complicated behavioral patterns (e.g. \cite{Freedman-2001}) as well as the impressive
efficiency (e.g. \cite{Laughlin-1998,Rieke-2000,Lewen-2001}) of nervous systems prove
them to be remarkable for information handling, for communication and as processing
systems. Information theory \cite{Shannon-1948,Cover-1991} provides a suitable
mathematical structure for quantifying properties of transmission lines. In particular,
it can be used to assess the amount of information that neural responses carry about the
outside world
\cite{Lewen-2001,Eckhorn-1974,Ruyter-1988,Theunissen-1996,Buracas-1998,Brenner-2000}. An
analog of this on the one--dimensional culture would be in measuring how much of
information injected at one end of the line actually makes it to the other end
\cite{Feinerman-2006}.

Information transmission rates through our one--dimensional systems have little
dependence on conduction speed. Rather, efficient communication relies on a substantial
variety of possible messages and a coding scheme that could minimize the effects of
transmission noise. Efficient information coding schemes in one-dimensional neuronal
networks are at the center of a heated debate in the neuroscience community. Patterned
neuronal cultures can be used to help verify and distinguish between the contradicting
models.

The Calcium imaging we used allows for the measurement of fluorescent amplitudes in
different areas across the culture. Amplitudes in single areas fluctuate with a typical
variation of $\pm 20\%$ between different bursts. The measured amplitudes are linearly
related to the population spiking rate (rate code), which averages the activity over
groups of about $100$ neurons and a time window of about $100$ milliseconds. This
experimental constraint is not essential and may be bypassed by using other forms of
recording (for example linear patterns on MEA dishes \cite{Jacobi-2006}), but it is
useful in allowing the direct evaluation of the stability of `rate coded' information as
it is transmitted across the culture.

The amplitude of the fluorescence signal progresses between
neighboring groups of neurons with unity gain. This means, for example, that if the
stimulated or `input' area produces a burst with fluorescence that is $10\%$ over its
average event, the neighboring area will tend to react the same $10\%$ over its average.
However, there is noise in the transmission. This noise becomes more dominant as the
`input' and `output' areas are further spaced. In fact, information can be transmitted
between two areas but decreases rapidly with the distance. Almost no information passes
between areas which are separated by more than ten average axonal lengths (about $4$ mm).

We modeled the decay of information along the one--dimensional culture by a Gaussian
relay channel \cite{Feinerman-2006}. In this model, the chain is composed of a series of
Gaussian information channels (with unity gain in this case) where the output of the
$n$th channel acts as input for the $(n+1)$th one. The information capacity of the
Gaussian chain is related to that of a single channel. This, in turn, depends only on its
signal to noise ratio (SNR).

To check this relation, the average SNR of a channel of a given length was measured in
the \invitro system and this value used to estimate the mutual information between areas
spaced at an arbitrary distance. The model produced an excellent quantitative fit to the
experimental results without any adjustable parameters. From this it was concluded that
rate coded information that is transmitted along the line decays only due to accumulating
transmission noise \cite{Feinerman-2006}.

It should be noted that classical correlation analysis is able to reveal only partially
the structures discovered by information theory. The reason is that information theory
allows an accurate comparison between information that is being carried in what may be
very different aspects of the same  signal (e.g., rate and temporal coding schemes as
discussed below) \cite{Hatsopoulos-1998}.

By externally stimulating uni--dimensional cultures at a single point we produced varied
responses that can be considered as different inputs into a transmission line.
Spontaneous activity also supplies a varied activity at one edge. The stimulated activity
then spreads across the culture by means of synaptic interaction to indirectly excite
distant parts of it (considered as outputs). At the `output', again, there is a variation
of possible responses. Predicting the `output' activity using the pre-knowledge of the
`input' activity is a means of evaluating the transmission reliability in the linear
culture and may be quantitatively evaluated by estimating the mutual information between
the activities of the two areas.

The activity of inhibitory neurons in the culture may be blocked by using
neurotransmitter receptor antagonists, leaving a fully excitatory network. Following this
treatment, event amplitudes as well as propagation speeds sharply increase. Information
transmission, on the other hand, goes down and almost no information passes even between
areas which are separated by just a single axonal length. This follows directly from the
fact that unity gain between areas is lost, probably because areas react with maximal
activity to any minimal input. Thus, we can conclude that the regulation provided by the
inhibitory network is necessary for maintaining the stability of the propagating signals.

\subsubsection{Further perspectives for temporal codes and rate codes}

A similar study was reported in two--dimensional cultures by Beggs {\it et al.}
\cite{Beggs-2003}. In that study it was concluded that transmission of rate code depends
on an excitatory/inhibitory balance that is present in the undisturbed culture. The
importance of such a balance for reliable information transmission was also discussed in
\cite{Beggs-2003,Vreeswijk-1996,Shadlen-1998}. Beggs {\it et al.} used two--dimensional
cultures and numerical simulations to measure distributions of the fraction of the total
number of neurons that participate in different events \cite{Beggs-2003}. They show that
this distribution exhibits a power law behavior typical of avalanche size distribution
that is theoretically predicted for a critical branching process. This criticality is
analogous to the unity gain measured on the one--dimensional culture. They suggest that
this implies that the natural state of the culture, in which information transmission is
maximized, is achieved by self organized criticality \cite{Bak-1998}.

There are two main hypotheses for cortical information representation
\cite{deCharms-1998,Rieke-1999}. The {\it independent coding hypothesis} suggests that
information is redundantly encoded in the response of single, independent neurons in a
larger population. Retrieval of reliable messages from noisy, error prone, neurons may be
achieved by averaging or pooling activity over a population or a time window. The {\it
coordinated coding hypothesis}, on the other hand, suggests that information is not
carried by single neurons. Rather, information is encoded, in the exact time lags between
individual spikes in groups of neurons.

Theoretical models of information transmission through linear arrays of neurons suggest
different coding solutions that fall into the framework of the two hypotheses presented
above. Shadlen {\it et al.} \cite{Shadlen-1998,Shadlen-1994} argue that the firing rate,
as averaged over a population of independent neurons over a time window of about $100$
milliseconds, is the efficient means of transmitting information through a linear
network. The averaging compensates for noise that is bound to accumulate as the signal
transverses through successive groups of neurons. This independent coding scheme is
referred to as {\it rate coding}. Some studies, however, convey the opposite view
\cite{Gautrais-1998,Diesmann-1999,Litvak-2003}. They show that as signals propagate along
a one--dimensional array of neurons, the spiking times of neighboring neurons tend to
synchronize and develop large scale correlations, entering what is called {\it synfire}
propagation \cite{Abeles-1991}. The advantages of averaging diminish as populations
become correlated and this causes firing rates to approach a predetermined fixed point as
the signal progresses. This fixed point has little dependence on the initial firing rate
so that all information coded in this rate is lost. The synchronization between neurons
does allow, however, the transmission of very precise spiking patterns through which
information could reliably be transmitted. This coordinated coding scheme is termed {\it
temporal coding} \cite{Rieke-1999}. The controversy between these two points of view,
which touches on the central question of how the brain represents information, is far
from being resolved.

Many information transmission models concentrate on uni--dimensional neuronal networks,
not only because of their simplicity, but also for their relation with living neural
networks. The cerebral cortex (the part of the brain responsible for transmitting and
processing sensory and motor information) is organized as a net of radial linear columns
\cite{Mountcastle-1957,Wilson-1973,Mountcastle-1997}. This fact supplies an important
motivation for generating 1--D models for information transmission
\cite{Litvak-2003,Wilson-1973}.

Another known \invivo linear system is the lateral spinothalamic pathway, which passes
rate coded sensory information from our fingers to our central nervous system (higher
rates correspond to stronger mechanical stimulation). This linear network uses very long
axons and only two synaptic jumps to relay the spike rate information. The necessity for
such an architecture can be explained by our results for the one--dimensional cultures
\cite{Feinerman-2006}. This is a clear case in which comparison to the 2D culture and the
living brain provides a new perspective: controlling connectivity can simplify networks
and also serve as an \invitro model of \invivo architectures. This perspective provides
novel possibilities by which cultured neurons can be used to study the brain.

\subsection{Parallel information channels}

Using our patterning techniques one can study more complex architectures to address
specific questions. Here we describe an experiment that asks about the information
transmission capacity of two parallel thin lines and its relation to their individual
capacities.

\begin{figure}
\begin{center}
\includegraphics[width=16cm]{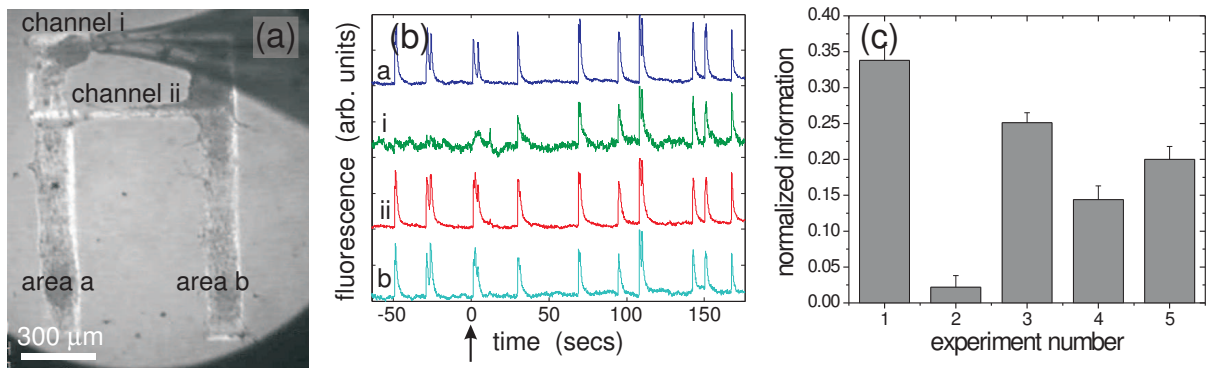}
\caption{Information capacity of two parallel channels. (a) Two thick areas ($170$ $\mu$m
wide) $a$ and $b$ are connected via two thin channels ($80$ $\mu$m wide), labeled $i$ and
$ii$. In this image, the concentric pipette, loaded with TTX, is in position to block
channel $i$, leaving channel $ii$ open. (b) Fluorescence levels of areas $a$ and $b$, and
channels $i$ and $ii$. TTX is applied until time $t=0$ (arrow). During TTX application,
channel $i$ is blocked (no signal), while areas $a$ and $b$, and channel $ii$ are
co--active. At $t>0$, TTX application is ceased and all four areas return to be
simultaneously active. (c) Mutual information between areas $a$ and $b$ normalized by
their average entropy \cite{Feinerman-2006}. The experiment number corresponds to: 1)
both channels open; 2) channel $i$ is blocked; 3) both channels open again; 4) channel
$ii$ is blocked; and 5) both channels open.}\label{Fig:information}
\end{center}
\end{figure}

Two thick areas connected by two thin 80 $\mu$m channels were patterned (Fig.\
\ref{Fig:information}a). The average mutual information between the two furthest areas on
the line (connected by two alternative routes one $2.6$ mm and the other $3.2$ mm long)
is $0.59 \pm 0.09$ bits as averaged over $386$ events in $11$ experiments. This is the
same as the information measured between two areas at the same distance connected by a
single $170$ $\mu$m thick line \cite{Feinerman-2006}. In the next part of the experiment,
each channel was separately blocked and reopened using concentric pipettes
\cite{Feinerman-2003} loaded with a spike preventing drug ($0.3$ $\mu$M of TTX) (Fig.\
\ref{Fig:information}b). The entropy in the two areas remained roughly the same (within
10\%) throughout the experiment, but the mutual information between them changed in
accordance with the application of TTX. As each channel is blocked the activity between
the areas remains coordinated (Fig.\ \ref{Fig:information}b) but mutual information
between the burst amplitudes goes down. Once TTX application is halted, the information
increases again (although to a value less than the original one by $20-25\%$, which may
be indicating that recovery after long periods of TTX application is not full) (Fig.\
\ref{Fig:information}c).

Our measurements on the parallel channels supplement the results of the previous
subsection. Since it is the spike rate (i.e., the amplitude of the signal) that carries
the information rather than the time between bursts, we can conclude that cooperation
between the channels is crucial for information transmission. Indeed, when both channels
are active, information transmission is enabled, with the same value as measured for a
single thick channel. This can be used to understand the nature of the code. For example,
specific temporal patterns would have been destroyed as the signal splits to two channels
due to the different traversal times and we can conclude that they are not  crucial (as
anticipated) for passing rate codes.

It would be interesting in the future to use this approach and architecture to
investigate other coding schemes.

\subsection{Speed of propagating fronts}

The next set of experiments deals with measuring causal progression of activity along
one--dimensional networks, and comparing it with models of wave propagation. Speed
measurements were preformed by Maeda {\it et al.} \cite{Maeda-1995} on semi
one--dimensional cultures, and by Feinerman {\it et al.} \cite{Feinerman-2005} for
one--dimensional ones.

The uni--dimensional culture displays population activity---monitored using the calcium
sensitive dye Fluo4---which commences at localized areas (either by stimulation or
spontaneously) and propagates to excite the full length of the line. This should be
contrasted with two--dimensional cultures, where front instabilities prevent a reliable
measurement of front propagation \cite{Kistler-2000,Wu-1999}.

The progressing activity front was tracked and two different regimes were identified.
Near the point of initiation, up to a few mean axonal lengths, the activity has low
amplitude and propagates slowly, at a speed of a few millimeters per second. Activity
then either decays or gains both in amplitude and speed, which rises up to $50-100$ mm/s,
and stably travels along the whole culture. Similar behavior has been observed in brain
slice preparations \cite{Haas-1984}. The speed of the high amplitude mode can be
controlled by modifying the synaptic coupling strength between the neurons\footnote{As a
standard procedure, synaptic strength is lowered by blocking the neuroreceptors with the
corresponding antagonists. In our experiments, AMPA--glutamate receptors in excitatory
neurons are blocked with the antagonist CNQX.} \cite{Feinerman-2005,Feinerman-2006}.

The appearance of two speeds can be understood in terms of an ``integrate and fire" (IF)
model \cite{Osan-2002,Osan-2004}. The IF model is a minimal model in which neurons are
represented by leaky capacitors \cite{Stein-1967}, which fire when their membrane
potential exceeds a certain threshold. The time evolution of the membrane potential
$u(t)$ is described by
\begin{equation}
\tau \frac{du(t)}{dt} = -u(t) + \sum_{\text{synapses}}\sum_{i=1}^{N}g_{\text{syn}}\;
\alpha\left(t-t_f^i\right),
\end{equation}
where $\tau$ is the membrane time constant. The sum above represents the currents
injected into a neuron, arriving from $N$ spikes. The synaptic strengths are labeled
$g_{\text{syn}}$, and $\alpha (t- t_f^i)$ represents the low--pass filtered synaptic
response to a presynaptic spike which fired at time $t_f^i$. As the voltage across the
capacitor exceeds a predefined `threshold potential', the IF neuron `spikes', an event
that is modeled as subsequent current injection into its postsynaptic neurons. The
voltage on the IF neuron is then reset to a voltage that is lower than the threshold
potential.

Continuous versions of the IF model were introduced in
\cite{Osan-2002,Osan-2004,Idiart-1993} by defining a potential $u(x,t)$ for every point
$x$ along a single dimension. The connectivity between neurons, $J(|y-x|)$, depends on
their distance only, and falls off with some relevant length scale (e.g., the average
axonal length). The model incorporates a single synapse type with time constant
$\tau_{\rm syn}$. The system is then described by
\begin{equation}
\tau \frac{\partial u(x,t)}{\partial t} = -u(x,t) + \int{J\left(|y-x|\right.)} \sum
g_{\text{syn}} \; \exp\bigl({-\frac{t-t_f^i}{\tau_{\text{syn}}}}\bigr)\;
H\left(t-t_f^i\right)\; dy~,
\end{equation}
where $H(t)$ is the Heaviside step function and the sum is, again, over all spikes in all
synapses.

In \cite{Feinerman-2005}, the model was simplified by assuming $N=1$. In this case the
wavefront may be fully analyzed in the linear regime that precedes the actual activity
front of spiking cells. This model can be analytically solved and predicts two speeds for
waves traveling across the system: a fast, stable wave and a slow, unstable one. As the
synaptic strength is weakened, the fast speed decreases drastically, while the slower one
gradually increases until there is breakup of propagation at the meeting point of the two
branches \cite{Feinerman-2005}. By introducing into the model the structural data
measured on the one--dimensional culture (neuronal density and axonal length
distribution), along with the well known single neuron time scales, one observes a strong
quantitative agreement between the \invitro experiment and the theoretical predictions
\cite{Feinerman-2005}.

The two wave speeds can be understood because there are two time scales in the model. The
fast wave corresponds to the fast time scale of the synaptic transmission, and is
relevant for inputs which coincide at this short scale of a few milliseconds. The slower
wave corresponds to the longer membrane leak time constant and involves input activity
which is less synchronized. A similar transition from asynchronous to synchronous
activity during propagation through a one--dimensional structure was previously predicted
by the theoretical model of Diesmann {\it et al.} \cite{Diesmann-1999}. This transition
plays an important role in information transmission capabilities as elaborated below.

\subsubsection{Perspectives and implications to other approaches}

\begin{figure}
\begin{center}
\includegraphics[width=7cm]{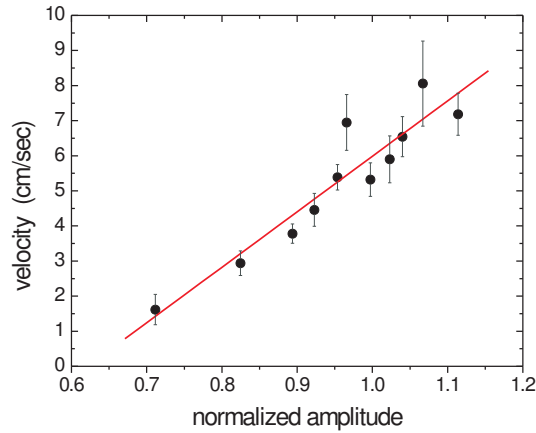}
\caption{Propagation speed as a function of burst amplitude. The plot summarizes $74$
spontaneous events that travel across a $17$ mm line. Amplitudes naturally vary between
bursts (about $\pm 20\%$ around the average normalized value of 1). The amplitude for
each event (while the signal travels along the line) is averaged over several areas on
the line (the two ends and the central part) \cite{Feinerman-2006}. Speed measurements
were preformed as described in \cite{Feinerman-2005} and are seen to scale linearly with
the amplitude.}\label{Fig:velocity} \end{center}
\end{figure}

Richness of the speed of the fast propagating front was also observed in some experiments
with brain slices \cite{Haas-1984,Bolea-2006}. We found that this fast speed scaled
linearly with the amplitude of the excitation, which measures the number of spikes in a
propagating burst (Fig.\ \ref{Fig:velocity}) (see also \cite{Jacobi-2006}). Larger
amplitudes signify more spiking in pre--synaptic cells and may be accounted for through a
proper re--scaling of the synaptic strength, $g_{\text{syn}}$. Such re--scaling should
take into account the relative synchronicity between spikes as only spikes that are
separated by an interval that is less than the membrane time constant can add up.
Multiple spike events are more difficult to understand theoretically. Osan {\it et
al}.~\cite{Osan-2004} expand their one--spike model into two spikes and demonstrate a
weak dependence of speeds on spike number. A more comprehensive analysis of traveling
waves that includes periodic spiking is introduced in \cite{Osan-2004,Bressloff-2000} and
predicts dispersion relations that may scale between $g_{\text{syn}}^{1/2}$ and
$g_{\text{syn}}^2$, so that a linear relation is not improbable. Golomb {\it et al.}
\cite{Golomb-1997} use a simulation of a one--dimensional network composed of
Hodgkin--Huxley type model neurons and two types of excitatory synapses (AMPA and NMDA)
to find a weak linear relation between propagation speed and the number of spikes in the
traveling front.

More complex behavior is predicted by theoretical models of one--dimensional neural
networks that incorporate mixed populations of excitatory and inhibitory neurons and
different connection footprints. Such networks were shown to support a variety of
propagating modes: slow--unstable, fast--stable, and lurching
\cite{Golomb-1999,Golomb-2001}. Similar models include bistable regimes where a
fast--stable and a partially slower--stable modes coexist and develop according to
specific initial conditions \cite{Golomb-2002,Compte-2003}.

\section{2-D neural cultures}
\label{sec:NEU-2D}

As we have seen, one--dimensional neural cultures allow us to study the speed of
propagating fronts, information content and coding schemes. In contrast, two--dimensional
neural cultures involve a large neural population in a very complex architecture.
Therefore, they are excellent model systems with which to study problems such as
connectivity and plasticity, which is the basis for learning and memory.

Abstract neural networks, their connectivity and learning have caught the attention of
physicists for many years \cite{Sompolinsky-1988,Sompolinsky-1990,Sompolinsky-1996}. The
advent of novel experimental techniques and the new awakening of graph theory---with its
application to a large variety of fields, from social sciences to genetics---allow one to
shed new light on the study of the connectivity in living neural networks.

The study of networks, namely graphs in which the nodes are neurons and the links are
connections between neurons, has received recently a vast theoretical input
\cite{Newman-2003,Boccaletti-2006,Newman-2006}. We are here not so much interested in the
scale--free graphs, although they seem omni--present in certain contexts, but rather in
the more information-theoretic aspects. These aspects have to do with what we believe
must be the main purpose of neurons: to communicate. An admittedly very simple instance
of such an approach is the study of the network of e-mail communications
\cite{Ebel-2002,EckmannMoses-2004}. In that case, encouraging results were found, and one
can try to transport the insights from those studies to the study of \invitro neural
networks. What they have in common with an e-mail network is the property that both
organize without any external master plan, obeying only constraints of space and time.

\subsection{Percolation in neural networks}\label{subsec:percolating_networks}

Neurons form complex webs of connections. Dendrites and axons extend, ramify, and form
synaptic links with their neighbors. This complex wiring diagram has caught the attention
of physicists and biologists for its resemblance to problems of percolation
\cite{Stauffer-1994,Bunde-1996}. Important questions are the critical distance that
dendrites and axons have to travel in order to make the network {\it percolate}, i.e., to
establish a path from one neuron of the network to any other, or the number of {\it
bonds} (connections) or {\it sites} (cell bodies) that can be removed without critically
damaging the functionality of the circuit. In the brain, neural networks display such
robust flexibility that circuits tolerate the destruction of many neurons or connections
while keeping the same, though degraded, function. For example, it is currently believed
that in Parkinson's disease, up to $70$\% of the functionality of the neurons in the
affected areas can be lost before behavioral symptoms appear \cite{Fearnley-1991}.

One approach that uses the concept of percolation combines experimental data of neural
shape and synaptic connectivity \cite{Costa-2003,Stepanyants-2005} with numerical
simulations to model the structure of living neural networks
\cite{Costa-2003,Costa-2005a,Costa-2005b}. These models are used to study network
dynamics, optimal neural circuitry, and the relation between connectivity and function.
Despite these efforts, an accurate mapping of real neural circuits, which often show a
hierarchical structure and clustered architecture (such as the mammalian cortex
\cite{Binzegger-2004,Sporns-2004}), is still unfeasible.

\subsection{Bond-percolation model}\label{subsec:percolation_model}

\begin{figure}
\begin{center}
\includegraphics[width=15cm]{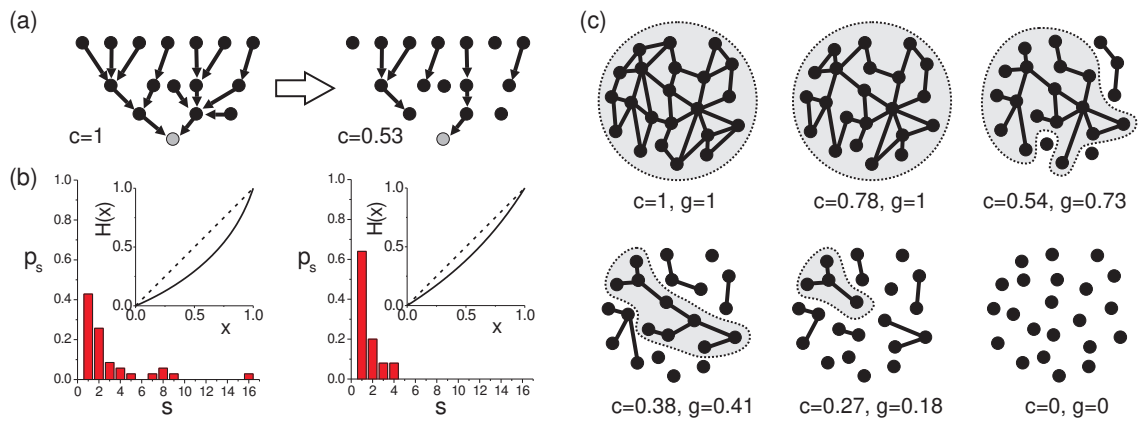}
\caption{(a) Percolation model. The neuron represented in grey fires either in response
to an external excitation or if any of its input neurons fire. At the highest
connectivity, this neuron has input clusters of size $s-1=0$ (the neuron responds to the
external excitation only), $7$ (left branch), $8$ (right branch), and $15$ (both
branches). At lower connectivity, its input--clusters are reduced to sizes $0$ and $3$.
(b) Corresponding $p_s$ distributions, obtained by counting all input clusters for all
neurons. Insets: The functions $H(x)$ (solid lines), compared with those for unconnected
neurons (dashed lines). (c) Concept of a giant component: The grey areas outline the size
of the giant component
 $g$ (biggest cluster) for gradually lower connectivity $c$.}\label{Fig:model}
\end{center}
\end{figure}

At the core of our experiments and model \cite{Breskin-2006} is a completely different
approach. We consider a simplified model of a neural network in terms of
bond--percolation on a graph. The neural network is represented by the directed graph $G$
with the following simplifying assumptions: \ A neuron has a probability $f$ to fire in
direct response to an external excitation (an applied electrical stimulus in the
experiments), and it always fires if any one of its input neurons fire (Fig.\
\ref{Fig:model}a).

The fraction of neurons in the network that fire for a given value of $f$ defines the
firing probability $\Phi (f)$. $\Phi (f)$ increases with the connectivity of $G$, because
any neuron along a directed path of inputs may fire and excite all the neurons downstream
(Fig.\ \ref{Fig:model}a). All the upstream neurons that can thus excite a certain neuron
define its input--cluster or excitation--basin. It is therefore convenient to express the
firing probability as the sum over the probabilities $p_{s}$ of a neuron to have an input
cluster of size $s-1$ (Fig.\ \ref{Fig:model}b),
\begin{eqnarray}
\Phi (f) &\nonumber=&f+(1-f)P\left( \mbox{any input neuron fires}\right)\\
&\label{Eq:Phi}=&f+(1-f)\sum_{s=1}^{\infty }p_{s}\left( 1-\left( 1-f\right) ^{s-
1}\right) =1-\sum_{s=1}^{\infty }p_{s}\left( 1-f\right) ^{s},
\end{eqnarray}
with $\sum\nolimits_{s}p_{s}=1$ (probability conservation). The firing probability $\Phi
(f)$ increases monotonically with $f$, and ranges between $\Phi (0)=0$ and $\Phi (1)=1$.
The connectivity of the network manifests itself by the deviation of $\Phi (f)$ from
linearity (for disconnected neurons one has $p_1=1$ and $\Phi (f)=f$). Equation
\eqref{Eq:Phi} indicates that the observed firing probability $\Phi (f)$ is actually one
minus the generating function $H(x)$ (or the $z$--transform) of the cluster--size
probability $p_{s}$ \cite{Harary-1952,Shante-1971},
\begin{equation}\label{Eq:Hx}
H(x)=\sum_{s=1}^{\infty }p_{s}x^{s}=1-\Phi (f)~,
\end{equation}
where $x=1-f$. One can extract from $H(x)$ the input--cluster size probabilities $p_{s}$,
formally by the inverse $z$--transform, or more practically, in the analysis of
experimental data, by fitting $H(x)$ to a polynomial in $x$.

In graph theory, say in a random graph, one considers connected components. When the
graph has $N$ nodes, one usually talks about a giant (connected) component if, in the
limit of $N\to\infty$, the largest component has a size which diverges with $N$
\cite{Borgs-2001}. Once a giant component emerges (Fig.\ \ref{Fig:model}c) the observed
firing pattern is significantly altered. In an infinite network, the giant component
always fires no matter how small the firing probability $f>0$ is. This is because even a
very small $f$ is sufficient to excite one of the infinitely many neurons that belong to
the giant component. This can be taken into account by splitting the neuron population
into a fraction $g$ that belongs to the giant component and always fires and the
remaining fraction $1-g$ that belongs to finite clusters (Fig.\ \ref{Fig:model}c). This
modifies the summation on cluster sizes into
\begin{eqnarray}
\Phi (f) &\nonumber=&g+(1-g)\left[f+(1-f)P\left( \mbox{any input neuron
fires}\right)\right]\\
&=&1-(1-g)\sum_{s=1}^{\infty }p_{s}\left( 1-f\right) ^{s}.
 \end{eqnarray}%
As expected, at the limit of almost no excitation $f\rightarrow 0$ only the giant
component fires, $\Phi (0)=g$, and $\Phi (f)$ monotonically increases to $\Phi (1)=1$.
With a giant component present the relation between $H(x)$ and the firing probability
changes, and Eq.\ \eqref{Eq:Hx} becomes
\begin{equation}\label{Eq:Hx-giant} H(x)=\sum_{s=1}^{\infty }p_{s}x^{s}=\frac{1-
\Phi (f)}{1-g}.
\end{equation}

As illustrated schematically in Fig.\ \ref{Fig:model}c, the size of the giant component
decreases with the connectivity $c$, defined as the fraction of remaining connections in
the network. At a critical connectivity $c_0$ the giant component disintegrates and its
size is comparable to the average cluster size in the network. This behavior suggests
that the connectivity undergoes a percolation transition, from a world of small,
disconnected clusters to a fast growing giant cluster that comprises most of the network.

The particular details of the percolation transition, i.e.,  the value of $c_0$ and how
fast the giant component increases with connectivity, depend on the degree distribution
of the neural network. Together with experiments and numerical simulations, as described
next, it is possible to use the percolation model to construct a physical picture of the
connectivity in the neural network.

\subsection{Joining theory and experiment}

\begin{figure}
\begin{center}
\includegraphics[width=12cm]{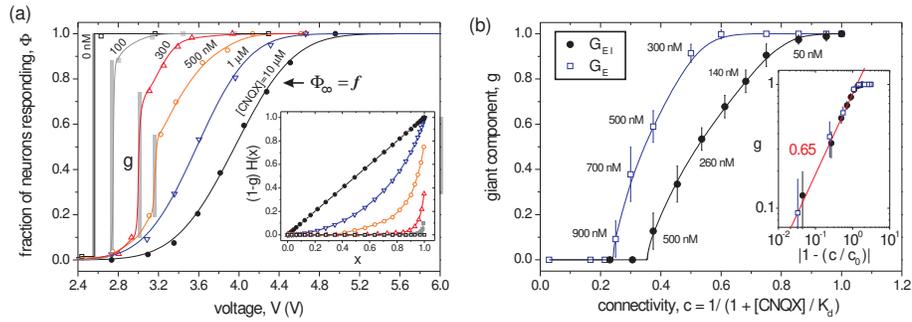} \vspace{0.0cm}
\caption{(a) Example of response curves $\Phi(V)$ for 6 concentrations of CNQX. The grey
vertical bars show the size of the giant component. They signal large jumps of the number
of neurons lighting up, for a small change of voltage. Thin lines are a guide to the eye
except for the $1$ $\mu$M and $10$ $\mu$M lines that are fits to error functions. Inset:
Corresponding $H(x)$ functions. The bar shows the size of the giant component for $300$
nM. (b) Size of the giant component as a function of the connectivity $c$, for the
network containing both excitatory and inhibitory neurons ($G_{EI}$, circles), and a
network with excitatory neurons only ($G_E$, squares). Lines are a guide to the eye. Some
CNQX concentrations are indicated for clarity. Inset: Log--log plot of the power law fits
$g \sim |1-c/c_o|^{\beta}$. The slope $0.65$ corresponds to the average value of $\beta$
for the two networks. Adapted from I. Breskin, J. Soriano, E. Moses, T. Tlusty, Phys.
Rev. Lett. 97, 188102, {\copyright} 2006 by the American Physical Society.}
\label{Fig:percolation}
\end{center}
\end{figure}

In our experiments \cite{Breskin-2006} we consider cultures of rat hippocampal neurons
plated on glass coverslips, and study the network response (fluorescence as described
earlier) to a collective electric stimulation. The network response $\Phi(V)$ is
quantified in terms of the fraction of neurons that respond to the external excitation at
voltage $V$. When the network is fully connected, the excitation of a small number of
neurons with low firing threshold suffices to light up the entire network. The response
curve is then similar to a step function, as shown in Fig.\ \ref{Fig:percolation}a.
Gradual weakening of the synaptic strength between neurons, which is achieved by blocking
the AMPA--glutamate receptors of excitatory neurons with the antagonist CNQX
\cite{Breskin-2006}, breaks the network off in small clusters, while a giant cluster
still contains most of the neurons. The response curves are then characterized by a
sudden jump that corresponds to the giant cluster (or giant component) $g$, and two tails
that correspond to clusters of neurons with either small or high firing threshold. At the
extreme of full blocking the network is completely disconnected, and the response curve
$\Phi_{\infty}(V)$ is characterized by the response of individual neurons.
$\Phi_{\infty}(V)$ is well described by an error function
$\Phi(V)=0.5+0.5\;\mathrm{erf}\left((V-V_0)/{\sqrt 2 \,\sigma_0}\right)$, indicating that
the firing threshold of individual neurons follows a Gaussian distribution with mean
$V_0$ and width $2\sigma_0$.

To study the size of the giant component as a function of the connectivity we consider
the parameter $c=1/(1+\text{[CNQX]}/K_d)$, where $K_d$ is the concentration of CNQX at
which $50$\% of the receptors are blocked \cite{Breskin-2006,Honore-1988}. Hence, $c$
quantifies the fraction of receptor molecules that are not bound by the antagonist CNQX
and therefore are free to activate the synapse. Thus, $c$ characterizes the connectivity
in the network, taking values between $0$ (full blocking) and $1$ (full connectivity).

The size of the giant component $g$ as a function of the connectivity is shown in Fig.\
\ref{Fig:percolation}b. Since neural cultures contain both excitatory and inhibitory
neurons, two kind of networks can be studied. $G_{EI}$ networks are those containing both
excitatory and inhibitory neurons. $G_E$ networks contain excitatory neurons only, with
the inhibitory neurons blocked with the antagonist bicuculine. The giant component in
both networks decreases with the loss of connectivity in the network, and disintegrates
at a critical connectivity $c_0$. We study this behavior as a percolation transition, and
describe it with the power law $g \sim |1 - c/c_0 |^{\beta}$ at the vicinity of the
critical point. Power law fits provide the same value of $\beta = 0.65 \pm 0.05$ within
experimental error (inset of Fig.~ \ref{Fig:percolation}b), suggesting that $\beta$ is an
intrinsic property of the network. The giant component for $G_{EI}$ networks breaks down
at a lower connectivity (higher concentration of CNQX) than for $G_E$ networks,
indicating that the role of inhibition is to effectively reduce the number of inputs that
a neuron receives on average.

The values of $c_0$ for $G_E$ and $G_{EI}$ networks, denoted by $c_e$ and $c_{ei}$
respectively, with $c_e \simeq 0.24$ and $c_{ei} \simeq 0.36$, provide an estimation of
the ratio between inhibition and excitation in the network. From percolation theory, $c_0
\sim 1/ n$, with $n$ the average number of connections per neuron. Thus, for $G_E$
networks, $c_e \sim 1 / n_e$, while for $G_{EI}$ networks $c_{ei} \sim 1 / (n_e - n_i)$
due to the presence of inhibition. The ratio between inhibition and excitation is then
given by $n_i / n_e = 1 - (c_e / c_{ei})$. This provides $75$\% excitation and $25$\%
inhibition in the neural culture, in agreement with the values reported in the
literature, which give an estimation of $70-80\%$ excitatory neurons
\cite{Marom-2002,Benson-1996}.

\begin{figure}
\begin{center}
\includegraphics[width=12cm]{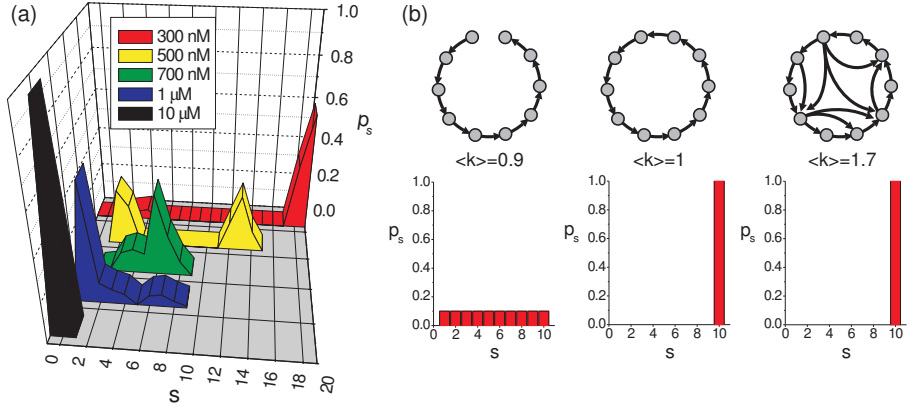} \vspace{0.0cm}
\caption{(a) Cluster size distribution $p_s$ for the generating functions shown in Fig.\
\ref{Fig:percolation}(a). The values in the table indicate the concentration of CNQX for
each curve. (b) Sensitivity of peaks in $p_s$ to loops. Left: neurons forming a
chain--like connectivity give a $p_s$ distributed uniformly. Center: closing the loop by
adding just one link collapses $p_s$ to a single peak. Right: additional links increase
the average connectivity $\langle k \rangle$, but do not modify $p_s$. Figure (a) adapted
from I. Breskin, J. Soriano, E. Moses, T. Tlusty, Phys. Rev. Lett. 97, 188102,
{\copyright} 2006 by the American Physical Society.} \label{Fig:Clusters}
\end{center}
\end{figure}

The response curves $\Phi(V)$ measured experimentally can be analyzed within the
framework of the model to extract information about the distribution of input clusters
that do not belong to the giant component. Since the response curve for a fully
disconnected network characterizes the firing probability $f(V)$ of independent neurons,
generating functions $H(x)$ can be constructed by plotting each response curve $\Phi(V)$
as a function of the response curve for independent neurons, $\Phi_{\infty}(V)$, as shown
in the inset of Fig.\ \ref{Fig:percolation}a. For those response curves with a giant
component present, its contribution is eliminated, and the resulting $H(x)$ function
re--scaled with the factor $1-g$, according to Eq.\ (\ref{Eq:Hx-giant}). The distribution
of input clusters $p_s$ is then obtained as the coefficients in the polynomial fits of
$H(x)$. Fig.\ \ref{Fig:Clusters}a shows the $p_s$ distribution for the $\Phi(V)$ curves
and $H(x)$ functions shown in Fig\ \ref{Fig:percolation}a.

The most remarkable feature of the $p_s$ distribution shown in Fig.\ \ref{Fig:Clusters}a
is that it is characterized by the presence of isolated peaks. These peaks are present
even for relatively high concentrations of CNQX (around $500$ nM). The explanation for
this observation is that the connectivity in the neural network is {\em not} tree--like,
but rather characterized by the presence of loops. As illustrated in Fig.\
\ref{Fig:Clusters}b, loops alter significantly the distribution of input clusters,
collapsing $p_s$ to single peaks. This idea is supported experimentally by the
observation that the areas of the neural culture with the highest density of neurons tend
to fire together in response to the external excitation. Their collective firing is
maintained even for high concentrations of CNQX, indicating that neurons may reinforce
their connections with close neighbors.

While the analysis of $H(x)$ provides a picture of the distribution of input clusters and
the connectivity at a local scale, the analysis of the exponent $\beta$ provides
information about the connectivity of the whole network. The exponent $\beta$, and the
behavior of the whole $g(c)$ curves shown in Fig.\ \ref{Fig:percolation}b, depend on the
actual distribution of connections per node in the network, $p_k$. Although the simple
percolation model can be extended to derive in principle $p_k$ from $H(x)$, the peaked
distribution of $p_s$ points at limitations of the analysis. Hence, numerical simulations
of the model have been explored to elucidate the degree distribution of the neural
network. Numerical simulations provide $\beta \simeq 0.66$ for a Gaussian degree
distribution, while they give $\beta \geq 1$ when the degree distribution follows a power
law \cite{Breskin-2006}, and this is clearly different from the experimental
observations. In should be kept in mind that the exponents we measure may differ from the
critical ones, since we measure them up to values of $c$ that are far from the transition
point.

Overall, we conclude that the neural culture can be viewed as a graph with a local
connectivity with Gaussian degree distribution and a significant presence of clusters or
loops that are maintained even at a low connectivity. Next, we will see how the analysis
of the percolation curves $g(c)$ can be used to extract additional information about the
connectivity in the neural network.

\subsection{Extracting biologically relevant properties from the percolation exponent}\label{subsec:Tsvi}

In this subsection, we will discuss how to combine mathematical constraints about
percolation with experimental observations to draw conclusions about the nature of the
network: degree distributions, length of neuronal connections, and the exponent $\beta $.


Two biologically relevant features of the living network arise in relation to the
percolation exponent of a random graph model. First, the living network is a directed
graph, so that we can make a distinction between the degree distribution of the {\em
input} graph (dendrites), for which the random graph model gives strong results, and the
{\em output} graph (axons), for which we can only make reasonable speculations.

Second, since the living network is located in real space, the links have a length
associated with them, while the random graph described by purely topological models does
not. We will see that the degree distribution is associated with a length distribution.

{\bf  The input graph degree distribution is Gaussian} We want to use the predictions of
the random graph model, and compare the measured growth curve of  the giant component to
what theory predicts for a general random graph \cite{Tlusty-2006}. We look at $p_{k}$,
the probability that a node has $k$ inputs, make the simplifying assumption that a node
fires if any of its inputs fires, and get the firing probability $\Phi (f)$ in terms of
$p_k$:
\begin{equation} \Phi (f)=f+(1-f)\sum_k p_k \left(1-\left(1-\Phi\right)^k\right). \end{equation}
with $f(V)$ again the probability of a single neuron to fire at excitation voltage $V$.
As usual, one forms the formal power series (or generating function)
\begin{equation} \widetilde{p}(z)\,=\,\sum_{s=1}^{\infty }p_{k}z^{k}~. \label{Eq:pk} \end{equation}
The giant component appears already at zero excitation for an infinite size network, we
therefore set $f=0$, which practically means that only the giant component lights up, so
that $\Phi=g$. From Eq.\ (\ref{Eq:pk}) we find that the probability for no firing, $1-g$,
is then a fixed point of the function $\widetilde{p}$ \begin{equation*}
1-g=\widetilde{p}(1-g). \end{equation*}
To proceed we need to know how the generating function is transformed when the edges are
diluted to a fraction $c$. It can be shown that it behaves according to
$\widetilde{p}(1-g)\rightarrow $ $\widetilde{p} (1-cg)$. Solving now the fixed point
equation
\begin{equation} 1-g\,=\,\widetilde{p}(1-cg) \end{equation}
for the fraction $g(c)$ of nodes in the giant component as a function of $c$, we get a
correspondence that depends on the edge distribution $p_{k}$.

The strength of this approach is that now the experimentally measured $g(c)$ can be
transformed to a result on $p_{k}$. In practice this necessitates some assumptions about
$p_{k}$, which can then be tested to see if they fit the experimental data. One popular
choice for $p_k$ is the scale-free graph, $p_{k}=\mathrm{const}\cdot k^{-\gamma }$. In
that case we get
\begin{equation} 1-g=\frac{\text{Li}_{\gamma }(1-cg)}{\zeta (\gamma )},  \label{sf} \end{equation}
where Li$_{\gamma}$ is the polylogarithmic function and $\zeta(\gamma)$ is the Riemann
zeta function. Another frequently studied distribution is the Erd\"os--R\'enyi (ER) one,
$p_{k} \sim\lambda ^{k}/k!$. In this case we obtain
\begin{equation} g(c)=1+\omega (-\lambda ce^{-\lambda c})/\lambda c,  \label{er} \end{equation}
 where Lambert's $\omega $ is defined as the solution of  $x=\omega (x)\exp[\omega (x)]$.
A comparison with the curves of Fig.\ \ref{Fig:percolation}b immediately shows that there
is a very good fit with Eq.\ (\ref{er}), while Eq.\ (\ref{sf}) shows poor agreement. One
can conclude from this that the input graph of neural networks is not scale-free.

{\bf Taking space into account} The graphs of living neural networks are realized in
physical space, in a two-dimensional environment very like a 2D lattice. What is absent
in the percolation theory is the notion of distances, or a {\em metric}. A seemingly
strong constraint therefore is the well-known and amply documented fact, that the
exponent $\beta $ in dimension $2$ with short-range connections is $5/36 \approx 0.14$
\cite{Stauffer-1994}. This is obviously very different from the measured value of about
$0.65$.

To study this difficulty, we first observe that long range connections can change the
exponent $\beta $. (For the moment, we ignore the fact that long range connections for
the input graph have been excluded, since we have shown that the input degree
distribution is Gaussian and the dendrites have relatively short length.) Take a lattice
in 2 dimensions, for example $\mathbf{Z}^{2}$ and consider the percolation problem on
this lattice. Then, there is some critical probability $p_{\mathrm{c}}$ with which links
have to be filled for percolation to occur; in fact $p_{\mathrm{c}}=1/2$. There are very
precise studies of the behavior for $p>p_{\mathrm{c}}$, and, as said above, in (large)
volume $N$ the giant component has size $N(p-p_{\mathrm{c}})^{5/36}$. What we propose is
to consider now a somewhat decorated lattice in which links of length $L$ occur with
probability proportional to $L^{-s}$, and we would like to determine $s$, knowing the
dimension $d=2$ and the exponent $\beta \simeq 0.65$. Like the links of the square
lattice, these long links are turned active or inactive with some probability. It is here
that we leave the domain of abstract graphs and consider graphs which are basically in
the plane. We conjecture that there is a relation between the dimension $d$, the decay
rate $s$ and the exponent $\beta$.  Results in this direction have been found long ago by
relating percolation to the study of Potts models in the limit of $q=1$. For example, it
is well-known, and intuitively easy to understand, that if there are too many long range
connections, the giant component will be of size $N$ and in fact this is reached for
$s\leq d$, see for example \cite{Berger-2002}, which contains also references to earlier
work. The variant of the question we have in mind does not seem to have been answered in
the literature. If there is indeed a relation  between $d$, $\beta$, and $s$, then it can
be used to determine $s$ from the experimentally measured values of $\beta$ and will give
information about the range of the neural connections.

We now come back to the neural network, keeping in mind that the input connections are
short, and have Gaussian distribution. We use the fact that the input and output graphs
of the living neural network are different to suggest a way to reconcile an observed
Gaussian distribution of input edges and the obvious existence of local connections, with
the small exponent of $5/36$ predicted for a locally connected graph in 2D. More
specifically, we use the fact that while axons may often travel large distances, the
dendritic tree is very limited. We assume that the number of output connections is
proportional to the length of the axon. Similarly the number of input connections is
proportional to the number of dendrites, to their length and to the density of axons in
the range of the dendritic tree.

A possible scenario for long-range connectivity relies therefore on a number of axons
that go a long distance $L$. Their abundance may occur with probability proportional to
$L^{-s}$, in which case the output graph will have a scale free distribution. However,
the input graph may still be Gaussian because there are a finite number of dendrites (of
bounded length) and each of them can only  accept a bounded number of axons.

Thus, we can conclude the following information about the neural network:

One limitation of using the exponent $\beta$ is the masking of the critical percolation
transition by finite size effects and by inherent noise. While the following analysis
gives excellent fits with the data, we have to remember that the biological system is not
ideal for extracting  $\beta $ accurately.

If $\beta $ is indeed the critical exponent, and the graph is not scale-free, then, a
change of $\beta $ from the 2-D value can perhaps be explained by assuming that a fixed
proportion of neurons extend a maximal distance. This structure will give the necessary
long-range connections, but its effect on the distribution of output connections will
only be to add a fixed peak at high connectivity. The edge distribution of the input
graph will not be affected by such a construction.

While the shortcomings of the discussion will be evident to the reader, we still hope
that this shows the important methodological conclusion that, by using theory and
experiment together, we are able to put strong constraints on the possible structures of
the living neural network.

\section{Acknowledgments}

We thank M. Segal, I. Breskin, S. Jacobi, and V. Greenberger for fruitful discussions and
technical assistance. J. Soriano acknowledges the financial support of the European
Training Network PHYNECS, project No. HPRN-CT-2002-00312. J.-P. Eckmann was partially
supported by the Fonds National Suisse and by the A.~Einstein Minerva Center for
Theoretical Physics. This work was also supported by the Israel Science Foundation, grant
No. 993/05, and the Minerva Foundation, Munich, Germany.



\end{document}